\documentclass[aps,prl,twocolumn,superscriptaddress,]{revtex4-1}
\usepackage{amsmath}
\usepackage{amsfonts}
\usepackage{graphicx}
\usepackage{color}
\usepackage{dsfont}
\usepackage{verbatim}
\usepackage{mathtools}
\usepackage[normalem]{ulem}
\usepackage{comment}
\usepackage{stackrel}
\usepackage{bm}
\usepackage[pdftex]{hyperref}
\hypersetup{colorlinks = true, urlcolor = black, linkcolor = black, citecolor = black}

\definecolor{myblue}{rgb}{.93, .93, 1}

\newcommand{\bsub}{\begin{subequations}}
	\newcommand{\esub}{\end{subequations}}

\newcommand{\vex}[1]{\bm{\mathrm{#1}}}

\begin{document}

	\title{Acoustic-phonon-mediated superconductivity in Bernal bilayer graphene}
	
	\author{Yang-Zhi~Chou}\email{yzchou@umd.edu}
	\affiliation{Condensed Matter Theory Center and Joint Quantum Institute, Department of Physics, University of Maryland, College Park, Maryland 20742, USA}

	\author{Fengcheng~Wu}
	\affiliation{School of Physics and Technology, Wuhan University, Wuhan  430072, China}
	
	\author{Jay D. Sau}
	\affiliation{Condensed Matter Theory Center and Joint Quantum Institute, Department of Physics, University of Maryland, College Park, Maryland 20742, USA}
	
	\author{Sankar Das~Sarma}
	\affiliation{Condensed Matter Theory Center and Joint Quantum Institute, Department of Physics, University of Maryland, College Park, Maryland 20742, USA}	
	\date{\today}
	
	\begin{abstract}
		We present a systematic theory of acoustic-phonon-mediated superconductivity, which incorporates Coulomb repulsion, explaining the recent experiment in Bernal bilayer graphene under a large displacement field. The acoustic-phonon mechanism predicts that $s$-wave spin-singlet and $f$-wave spin-triplet pairings are degenerate and dominant. Assuming a spin-polarized valley-unpolarized normal state, we obtain $f$-wave spin-triplet superconductivity with $T_c\sim 20$ mK near $n_e=-0.6\times 10^{12}$ cm$^{-2}$ for hole doping, in approximate agreement with the experiment. We further predict the existence of superconductivity for larger doping in both electron-doped and hole-doped regimes.
		Our results indicate that the observed spin-triplet superconductivity in Bernal bilayer graphene arises from acoustic phonons.
	\end{abstract}
	
	\maketitle
	
	\textit{Introduction.---}  
	New experiments in ABC-stacked rhombohedral trilayer graphene (RTG) \cite{Zhou2021,Zhou2021_SC_RTG} reported the existence of superconductivity and multiple symmetry-breaking phases, reminiscent of the phenomenology in moir\'e graphene systems \cite{Cao2018_tbg1,Cao2018_tbg2,Yankowitz2019,Polshyn2019,Cao2020PRL,Sharpe2019,Lu2019,Kerelsky2019,Jiang2019,Xie2019_spectroscopic,Choi2019,Serlin2020,Park2021flavour,Chen2019signatures,Burg2019,Shen2020correlated,Cao2020tunable,Liu2020tunable,Park2021tunable,Hao2021electric,Cao2021}. Remarkably, RTG allows for two distinct superconducting phases, corresponding to spin-singlet and spin-triplet pairings \cite{Zhou2021_SC_RTG}. The discovery of superconductivity in non-moir\'e RTG is a fundamental breakthrough as it implies that moir\'e and, by implication, strong correlations are not essential conditions for superconductivity in graphene-based systems.
	Shortly after the discovery of superconductivity in RTG, an electron-acoustic-phonon pairing mechanism was proposed as a likely explanation \cite{Chou2021_RTG_SC}. Alternative theoretical explanations focusing on electron-electron interactions were also proposed afterward \cite{Chatterjee2021,Ghazaryan2021,Dong2021,Cea2021,Szabo2021,You2021}.
	
\begin{figure}[t!]
		\includegraphics[width=0.425\textwidth]{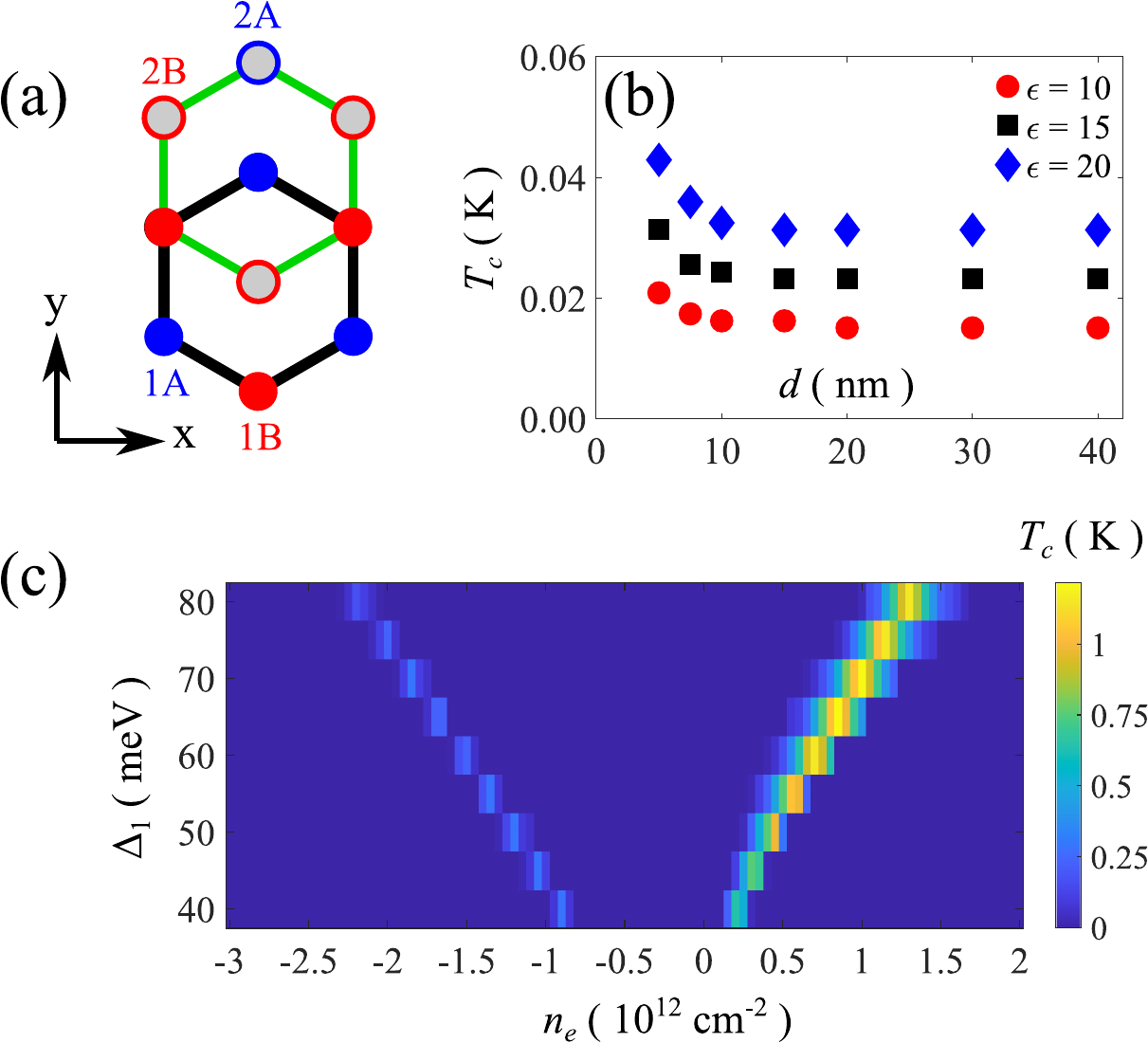}
		\caption{Summary of the main results. (a) Top view of BBG. (b) Predicted $T_c$ of spin-triplet SC and the dependence on half gate distance ($d$) and dielectric constant ($\epsilon$). We assume a spin-polarized normal state with $\Delta_1=50$ meV and $n_e=-0.6\times 10^{12}$ cm$^{-2}$. The results are consistent with the experimentally extracted $T_c$ for spin-triplet SC in Ref. \cite{Zhou2021_BBG}. Enhancement of superconductivity by tuning gate screening is a crucial manifestation of our theory, which can be tested experimentally.
			(c) Calculated $T_c$ as a function of $n_e$ and $\Delta_1$ for an unpolarized normal state. }
		\label{Fig:BBG}
	\end{figure}

	RTG is not the only superconducting possibility in the moir\'eless graphene-based systems.
	A very recent experiment \cite{Zhou2021_BBG} demonstrated that superconductivity can emerge in Bernal bilayer graphene (BBG), Fig.~\ref{Fig:BBG}(a), after applying an in-plane magnetic field that suppresses a competing ordered state. The highest reported superconducting temperature in BBG is around 30 mK, and superconductivity persists for an in-plane magnetic field $B_{\parallel}\approx 150$ mT, implying a likely spin-triplet pairing. The experiment in BBG revealed a number of features that are qualitatively similar to RTG. Therefore, it is reasonable to ask if there is a universal phonon-mediated superconducting mechanism in graphene-based materials, particularly since even the observed superconductivity in twisted bilayer graphene appears consistent with the acoustic phonon mechanism \cite{Wu2019_phonon}. 
	
	In this Letter, we develop a systematic theory of acoustic-phonon-mediated BBG superconductivity incorporating a realistic band structure and the detrimental effect of Coulomb repulsion. We show that acoustic-phonon-mediated superconductivity can indeed arise in BBG for a wide range of doping, and $s$-wave spin-singlet and $f$-wave spin-triplet pairings are degenerate and dominant. Assuming a spin-polarized normal state (because of the applied magnetic field), we obtain $f$-wave spin-triplet superconductivity with a calculated $T_c\sim 20$ mK near hole-doping $n_e=-0.6\times 10^{12}$ cm$^{-2}$ as shown in Fig.~\ref{Fig:BBG}(b). We obtain the gate distance ($2d$) and dielectric constant ($\epsilon$) dependence of $T_c$ based on our microscopic phonon theory.
	Our results suggest that the spin-triplet superconductivity in BBG is likely due to acoustic phonons. We predict the existence of superconductivity for the higher doping region not yet explored experimentally as shown in Fig.~\ref{Fig:BBG}(c).

	\textit{Single-particle model.---} The BBG is made of two layers of graphene sheets that are stacked such that the B sites of the top layer are on the A sites of the bottom layer as plotted in Fig.~\ref{Fig:BBG}(a).
	The low-energy single-particle states near the $K$ and $-K$ valleys can be described by a $\vex{k}\cdot\vex{p}$ Hamiltonian \cite{Jung2014} given by
	\begin{align}\label{Eq:H_0}
		\hat{H}_0=\sum_{\vex{k}}\hat{\Psi}^{\dagger}(\vex{k})\hat{h}_{\vex{k}}\hat{\Psi}(\vex{k}),
	\end{align}
	where $\hat{h}(\vex{k})=\left[\hat{h}_+(\vex{k})\oplus \hat{h}_-(\vex{k})\right]\hat{1}_s$, $\hat{h}_{\pm}(\vex{k})$ is a $4\times 4$ matrix encoding the low-energy bands near the $\pm K$ valley, $\hat{1}_s$ is the identity matrix in the spin space, and $\hat{\Psi}(\vex{k})$ is a 16-component column vector consisting of an electron annihilation operator $\psi_{\tau\sigma l s}$ with valley $\tau$, sublattice $\sigma$, layer $l$, and spin $s$. Note that the momentum $\vex{k}$ in Eq.~(\ref{Eq:H_0}) is relative to the $K$ or $-K$ point.
	Our theory of BBG superconductivity is explicitly based on the band structure model of Eq.~(\ref{Eq:H_0}) using realistic band parameters \cite{Jung2014}.
	
	The low-energy bands of BBG have large probability on the A sites of the top layer (1A) and B sites of the bottom layer (2B). The sublattice polarization generically suppresses electron-optical-phonon coupling \cite{Choi2021dichotomy,Wu2018,Wu2020_TDBG}, making electron-acoustic-phonon coupling the dominating contribution. Similar to RTG \cite{Chou2021_RTG_SC,Zhang2010}, the superconducting states with intralayer intersublattice pairings should be generically suppressed for low-energy BBG bands because one of the sublattices in each layer has higher energy.
	
	We formally diagonalize the Hamiltonian in Eq.~(\ref{Eq:H_0}) as follows:
	\begin{align}\label{Eq:H_0_diagonalized}
		\hat{H}_0=\sum_{\tau=\pm}\sum_{b=1}^4\sum_{s=\uparrow,\downarrow}\mathcal{E}_{\tau,b}(\vex{k})c^{\dagger}_{\tau b s}(\vex{k})c_{\tau b s}(\vex{k}),
	\end{align}
	where $\mathcal{E}_{\tau,b}(\vex{k})$ encodes the energy-momentum dispersion of the $b$th band and valley $\tau K$, and $c_{\tau b s}(\vex{k})$ is an electron annihilation operator of valley $\tau K$, the $b$th band, spin $s$, and momentum $\vex{k}$. The operators $\psi_{\tau\sigma l s}$ (in microscopic basis) and $c_{\tau b s}$ (in band basis) obey $\psi_{\tau\sigma l s}(\vex{k})=\sum_b\Phi_{\tau b, \sigma l}(\vex{k})c_{\tau b s}(\vex{k})$, where $\Phi_{\tau b, \sigma l}(\vex{k})$ is the wavefunction of valley $\tau$K and band $b$. The (spinless) time-reversal symmetry yields further constraints: $\mathcal{E}_{+,b}(\vex{k})=\mathcal{E}_{-,b}(-\vex{k})$ and $\Phi_{+ b, \sigma l}(\vex{k})=\Phi_{- b, \sigma l}^*(-\vex{k})$. 
	
\begin{figure}[t!]
		\includegraphics[width=0.4\textwidth]{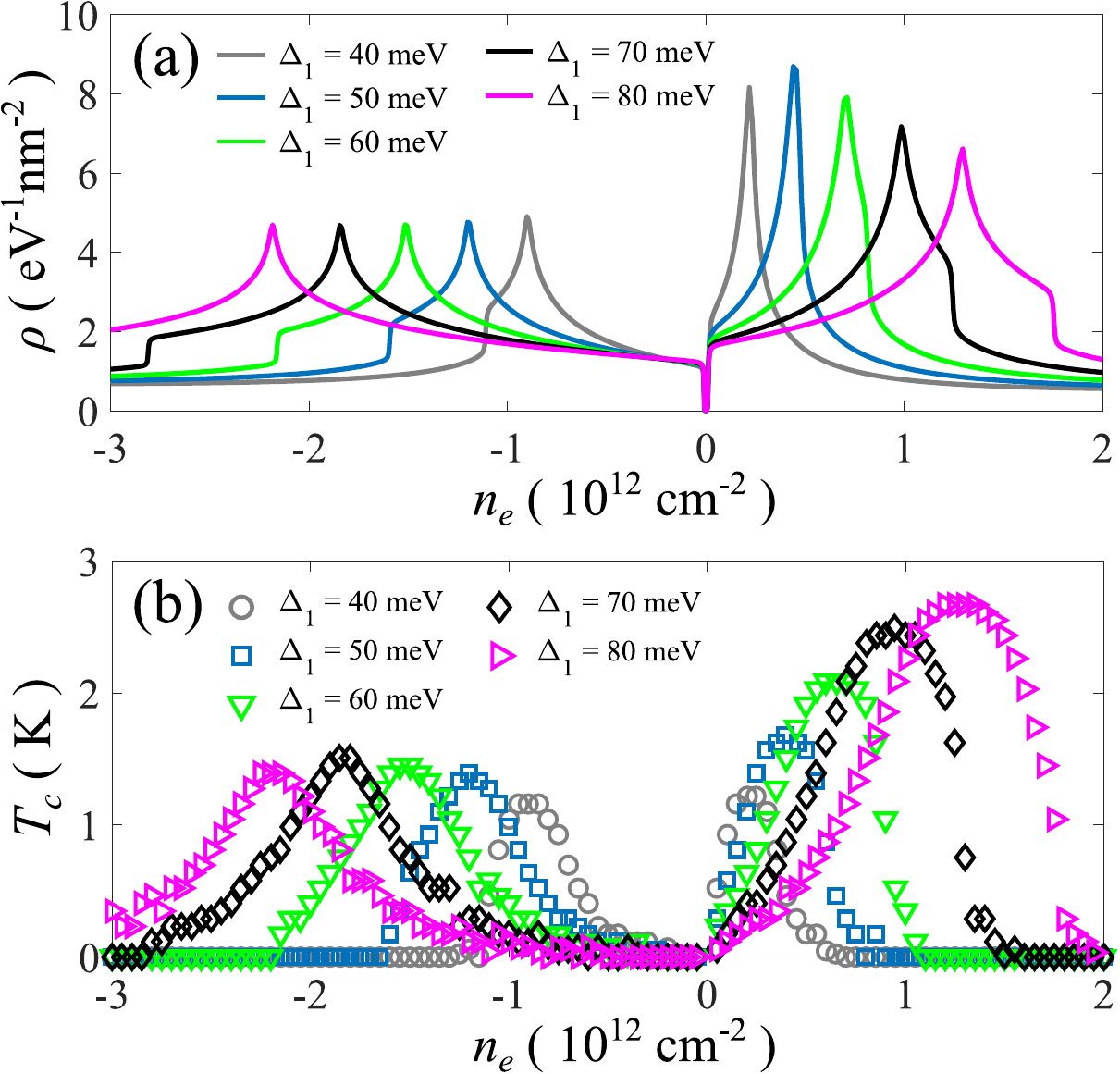}
		\caption{Total density of states ($\rho$) and superconducting transition temperature ($T_c$) without Coulomb repulsion. $n_e$ is the total doping density including spin and valley degeneracy. (a) $\rho$ as a function of $n_e$ for $\Delta_1=40-80$ meV. (b) $T_c$ for acoustic-phonon-mediated superconductivity (without Coulomb repulsion) as a function of $n_e$ for $\Delta_1=40-80$ meV. To extract $T_c$, we solve Eqs.~(\ref{Eq:LGE_1}) and (\ref{Eq:LGE_2}) with 5000 energy levels from a fine momentum grid with a spacing $\Delta k=0.0021a_0^{-1}$. $a_0=0.246$ nm is the lattice constant of graphene.  }
		\label{Fig:DOS_Tc}
	\end{figure}

	We model the displacement field (i.e., the applied out-of-plane electric field) by adding $-\Delta_1$ and $\Delta_1$ to the top layer and the bottom layer respectively. See Supplemental Material (SM) for discussions on the Fermi surface and the Van Hove singularity (VHS) \cite{SM}. With a nonzero $\Delta_1$, a band gap develops at the charge neutrality point, and the low-energy bands manifest large density of states (DOS) in both conduction and valence bands. We focus on the regime with $|\Delta_1|> 40$ meV \cite{D1} where large DOS is developed over a wide range of carrier densities, consequently enhancing the superconducting possibility. 
	In Fig.~\ref{Fig:DOS_Tc}(a), we plot the DOS as a function of carrier density. The above-mentioned band properties are qualitatively reminiscent of RTG, but the DOS is smaller in BBG (and hence generally weaker in superconductivity).
	
	\textit{Acoustic-phonon-mediated superconductivity.---} We study superconductivity mediated by the in-plane longitudinal acoustic phonons \cite{Wu2019_phonon,Wu2020_TDBG}. The BCS pairing interaction can be derived by ignoring the frequency dependence in the phonon propagator, and the effective attractive interaction is given by
	\begin{align}\label{Eq:H_BCS}
		\hat{H}_{\text{BCS}}=-g_0\sum_{\sigma,\sigma',l,s,s'}\int d^{2}\vex{r}\,\psi^{\dagger}_{+\sigma l s}\psi^{\dagger}_{-\sigma' l s'}\psi_{-\sigma' l s'}\psi_{+\sigma l s},
	\end{align}
	where $g_0$ is the coupling constant of the acoustic-phonon-mediated attraction. The interaction in Eq.~(\ref{Eq:H_BCS}) is intralayer because it is inherited from in-plane longitudinal acoustic phonons. The coupling constant $g_0=D^2/(\rho_m v_s^2)$, where $D$ is the deformation potential, $\rho_m$ is the mass density of monolayer graphene, and $v_s$ is the sound velocity. Using $D=30$ eV \cite{DeformationP}, $\rho_m=7.6\times 10^{-8}$ g/cm$^2$ \cite{Efetov2010,Hwang2008}, and $v_s=2\times 10^6$ cm/s, we obtain $g_0\approx474$ meV$\cdot$nm$^2$ \cite{Wu2020_TDBG,Chou2021_RTG_SC}. Notice that the magnitude of $D$ is not precisely known, and it might be off by a factor of $2$ \cite{Wu2019_phonon,Hwang2008}.
	
	The validity of BCS approximation relies on the retardation effect, i.e., phonons are slower than electrons. However, this might not be true for systems strongly affected by a nearby VHS where the Fermi velocity may be suppressed. To check this, we estimate average the Fermi velocity, $\bar{v}_F=2\sqrt{|n_e|}/(\hbar\sqrt{\pi}\rho)$, where $n_e$ is the carrier density and $\rho$ is the total DOS (incorporating spin and valley).
	For generic doping densities, we find that $\bar{v}_F$ is larger than the sound velocity $v_s$, suggesting the validity of the Migdal theorem and BCS approximation in BBG. See SM for $\bar{v}_F$ as a function of $n_e$ \cite{SM}.
	
	We now discuss the pairing symmetry in the low-energy BBG bands. Only the intervalley Cooper pairs are considered here because $\mathcal{E}_{\tau,b}(\vex{k})\neq \mathcal{E}_{\tau,b}(-\vex{k})$ generically suppresses the intravalley superconductivity \cite{Einenkel2011,Sun2021}. Following the classification scheme in literature \cite{Wu2019_phonon,Wu2020_TDBG,Chou2021correlation,Chou2021_RTG_SC}, the intervalley pairing symmetry (i.e., $s$-, $p$-, $d$-, and $f$- wave) can be determined from $\mathcal{C}_{3z}$ (threefold rotation about the hexagon center) and spin SU(2) symmetry. Similar to RTG \cite{Chou2021_RTG_SC}, we find that the  intralayer intersublattice pairings are strongly suppressed in the low-energy bands as one of the sublattices in each layer is at high energy. Therefore, we focus only on the intralayer intrasublattice pairings, i.e., $s$-wave spin-singlet and $f$-wave spin-triplet pairings, in the rest of this work.

	To simplify the calculations, we adopt the single-band approximation to where the Fermi energy $\mathcal{E}_F$ lies. This approximation is valid because of the energy separation to remote bands. As we discussed previously, only the intervalley intrasublattice pairings are taken into account. The projected BCS pairing (to the $b$th band) is given by
	\begin{align}
		\label{Eq:H_BCS_b}\hat{H}_{\text{BCS}}'=&\frac{-1}{\mathcal{A}}\sum g_{\vex{k},\vex{k}'}^{(b)}c^{\dagger}_{+bs}(\vex{k})c^{\dagger}_{-bs'}(-\vex{k})c_{-bs'}(-\vex{k}')c_{+bs}(\vex{k}'),\\
		\label{Eq:g_kk'}g_{\vex{k},\vex{k}'}^{(b)}=&g_0\sum_{\sigma,l}\left|\Phi_{+b;l\sigma}(\vex{k})\right|^2\left|\Phi_{+b;l\sigma}(\vex{k}')\right|^2,
	\end{align}
	where $\sum\equiv \sum_{s,s'}\sum_{\vex{k},\vex{k'}}$ in Eq.~(\ref{Eq:H_BCS_b}), $\mathcal{A}$ is the area of the system, and $g_{\vex{k}.\vex{k}'}^{(b)}$ is the momentum-dependent coupling constant in the $b$th band. Because the acoustic-phonon-mediated attraction respects an enlarged SU(2)$\times$SU(2) symmetry [i.e., independent spin rotational SU(2) symmetry within each valley], the $s$-wave spin-singlet and the $f$-wave spin-triplet pairings are exactly degenerate. The degeneracy can be broken by either the subleading pairing glues (e.g., optical phonons which enhances the $s$-wave spin-singlet pairing) or by symmetry breaking perturbations (e.g., an applied Zeeman field which favors the $f$-wave spin-triplet pairing).
	
	Within the mean-field approximation, one can derive the linearized gap equation as follows (see derivations in SM \cite{SM}):
	\begin{align}
		\label{Eq:LGE_1}\Delta_{s's}(\vex{k})=&\sum_{\vex{k}'}\chi_{\vex{k},\vex{k}'}\Delta_{s's}(\vex{k}'),\\
		\label{Eq:LGE_2}\chi_{\vex{k},\vex{k}'}=&\frac{g_{\vex{k},\vex{k}'}^{(b)}}{\mathcal{A}}\frac{\tanh\left[\frac{\mathcal{E}_{+b}(\vex{k}')-\mathcal{E}_F}{2k_BT}\right]}{2\mathcal{E}_{+b}(\vex{k}')-2\mathcal{E}_F},
	\end{align}
	where $k_B$ is the Boltzmann constant, $\mathcal{E}_F$ is the Fermi energy, and the superconducting order parameter is defined by
	\begin{align}
		\label{Eq:Delta}\Delta_{s's}(\vex{k}')=&\frac{1}{\mathcal{A}}\sum_b\sum_{\vex{k}'}g_{\vex{k},\vex{k}'}^{(b)}\left\langle c_{-bs'}(-\vex{k}')c_{+bs}(\vex{k}')\right\rangle.
	\end{align}
	The transition temperature $T_c$ is determined by the highest $T$ such that $\chi_{\vex{k},\vex{k}'}$ yields an eigenvalue $1$. 
	
	We numerically solve Eqs.~(\ref{Eq:LGE_1}) and (\ref{Eq:LGE_2}) for $\Delta_1=40-80$ meV and extract $T_c$ as a function of doping in Fig.~\ref{Fig:DOS_Tc}(b). In Fig.~\ref{Fig:DOS_Tc}(b), the highest $T_c$ is around 2.5 K near VHS, and $T_c$ remains finite in a wide range of doping. Similar to RTG \cite{Chou2021_RTG_SC,Lothman2017}, the prevalence of superconductivity arises from the energy dependence of $\chi_{\vex{k},\vex{k}'}$ [Eq.~(\ref{Eq:LGE_2})], which allows contributions from states away from $\mathcal{E}_F$. Due to the SU(2)$\times$SU(2) symmetry of the acoustic-phonon-mediated attraction, the $s$-wave spin-singlet and $f$-wave spin-triplet are exactly degenerate. Our results indicate that the electron-acoustic-phonon mechanism is the likely explanation for superconductivity in BBG, similar to the previous results for RTG \cite{Chou2021_RTG_SC}, although our calculated $T_c$ in Fig. \ref{Fig:DOS_Tc}(b) is relatively high (of course, lowering the coupling $g_0$ by using a smaller value of the deformation potential constant would suppress $T_c$ exponentially).
	
	\textit{Suppression from Coulomb repulsion.---} So far, we use the BCS interaction in Eq.~(\ref{Eq:H_BCS}) to estimate $T_c$. Coulomb repulsion can reduce the effective attraction and suppress superconductivity. To investigate if superconductivity survives in the presence of Coulomb repulsion, we qualitatively derive the effective pairing interaction incorporating the so-called $\mu^*$ effect \cite{MorelAnderson1962} in the following.
	
	In two-dimensional materials, Coulomb interaction is screened by the dielectric environment and nearby metallic gates. We assume, consistent with experiments, that BBG is capped by an insulator and is in the middle of two metallic gates, and then the screened Coulomb interaction is given by \cite{Wu2020_TDBG}:
	\begin{align}\label{Eq:V_C}
		V_{\text{C}}(\vex{q})=\frac{2\pi e^2}{\epsilon}\frac{\tanh(|\vex{q}|d)}{|\vex{q}|}.
	\end{align}
	In addition, the intraband screening by the carriers themselves might be significant due to the large DOS in BBG.
	To incorporate the intraband screening, we adopt the Thomas-Fermi approximation given by
	\begin{align}\label{Eq:V_TF}
		V_{\text{TF}}(\vex{q};\mathcal{E}_F)=\frac{V_{\text{C}}(\vex{q})}{1+V_{\text{C}}(\vex{q})\rho(\mathcal{E}_F)},
	\end{align}
	where $\rho(\mathcal{E}_F)$ is the total DOS at Fermi energy. The Thomas-Fermi approximation is the static limit of the random phase approximation, obtained by resumming over bubble diagrams. When $V_{\text{C}}(\vex{q})\rho(\mathcal{E}_F)\gg 1$, $V_{\text{TF}}(\vex{q};\mathcal{E}_F)\approx 1/\rho(\mathcal{E}_F)$, which is independent of $\epsilon$ and $d$. 
	
	\begin{figure}[t!]
		\includegraphics[width=0.35\textwidth]{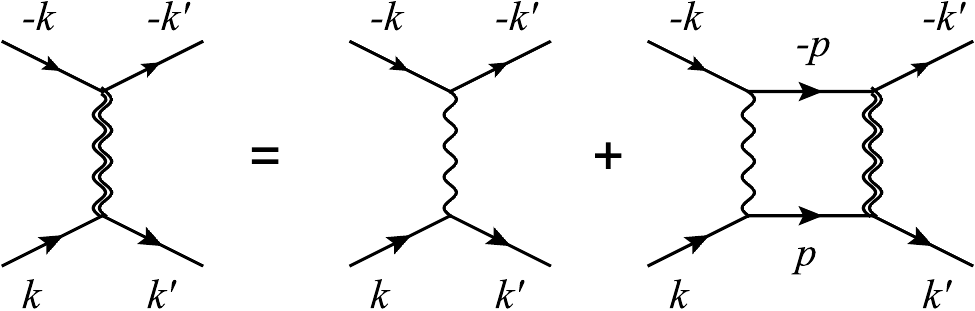}
		\caption{Diagrammatic representation of self-consistent ladder equation for the $\mu^*$ effect. The single wiggly lines denote the bare interaction $U_0$; the double wiggly lines denote the renormalized interaction $U_R$; the solid lines with arrows denote the electron propagators. Note that $k$, $k'$, and $p$ are in valley $K$ while $-k$, $-k'$, and $-p$ are in valley $-K$.
		}
		\label{Fig:Ladder}
	\end{figure}

	Besides screening, the renormalization from the high energy states can also reduce the Coulomb repulsion in the BCS channel, and we treat this by solving the ladder self-consistent equation \cite{Coleman2015introduction} shown in Fig.~\ref{Fig:Ladder}. If we ignore the momentum dependence of $V_{\text{TF}}(\vex{q};\mathcal{E}_F)$ and use $U_0(\mathcal{E}_F)\equiv V_{\text{TF}}(k_F;\mathcal{E}_F)$, the renormalized interaction is given by
	\begin{align}\label{Eq:U_R}
		U_R(\mathcal{E}_F)=\frac{U_0(\mathcal{E}_F)}{1+U_0(\mathcal{E}_F)\Gamma(\mathcal{E}_F;\omega_c;\Lambda)},
	\end{align}
	where $\Gamma(\mathcal{E}_F;\omega_c;\Lambda)$ is a function encoding the renormalization from the energies satisfying $\omega_c<|\mathcal{E}_{\tau b}(\vex{k})-\mathcal{E}_F|<\Lambda$, $\omega_c=2v_sk_F$, and $\Lambda$ is the energy cutoff. Equation~(\ref{Eq:U_R}) denotes the effective Coulomb repulsion within an energy range $[\mathcal{E}_F-\omega_c,\mathcal{E}_F+\omega_c]$. See SM for a derivation of $\Gamma$ \cite{SM}. 
	
\begin{figure}[t!]
		\includegraphics[width=0.375\textwidth]{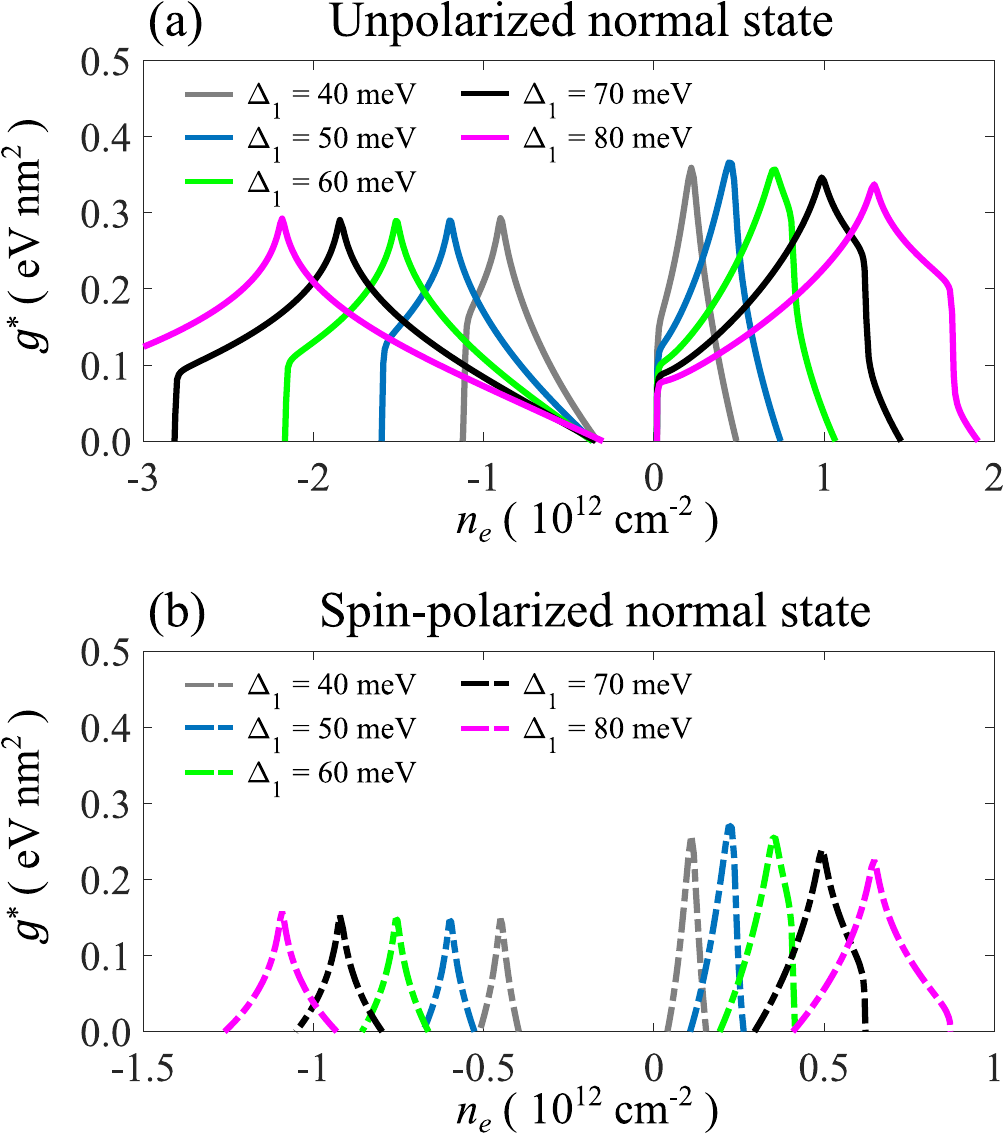}
		\caption{Effective attraction ($g^*$) in (a) an unpolarized metal and (b) a spin-polarized metal. Only attractive $g^*$ is shown. We use the half gate distance $d=40$ nm and dielectric constant $\epsilon=10$ here. Note that the scales in $n_e$ are different by a factor of 2 between (a) and (b) due to spin polarization in (b). The peaks in each curve correspond to VHS. $g^*$ is generically smaller in the spin-polarized case due to $\rho(\mathcal{E}_F)$ being half of that in the unpolarized case. 
		}
		\label{Fig:MuStar}
	\end{figure}

	We first concentrate on an unpolarized normal state (i.e., absence of flavor polarization) and plot the associated effective pairing $g^*=g_0-U_R$ as a function of doping for $\Delta_1=40-80$ meV in Fig.~\ref{Fig:MuStar}(a). Remarkably, $g^*$ remains positive for a wide range of carrier density, suggesting that superconductivity persists in the presence of Coulomb repulsion. 
	Replacing $g_0$ by $g^*$ in Eq.~(\ref{Eq:g_kk'}) and solving Eqs.~(\ref{Eq:LGE_1}) and (\ref{Eq:LGE_2}), we confirm that superconductivity indeed survives the Coulomb repulsion, but $T_c$ is noticeably reduced because the effective Coulomb repulsion suppresses the phonon-mediated attraction causing the pairing. 
	The resulting $T_c$ is insensitive to $\epsilon$ and $d$, indicating a strong screening situation.
	In Fig.~\ref{Fig:BBG}(c), $T_c$ as a function of $n_e$ and $\Delta_1$ is summarized, providing a detailed map for the search of superconductivity. The highest $T_c$ is around 0.3 K (1.2 K) in the hole (electron) doping. The actual $T_c$ would be lower since our mean-field BCS theory in general overestimates $T_c$ in two dimensions. Close to VHS (at which $g^*$ is peaked), the non-adiabatic vertex corrections \cite{Phan_2020,Cappelluti1996}, which we ignore, can also modify $T_c$.

	For spin-polarized valley-unpolarized normal states, $\rho(\mathcal{E}_F)$ is half of the value in an unpolarized state, so the intraband screening is weaker, resulting in a larger reduction of effective attraction. Therefore, we generically predict that the superconductivity arising in a spin-polarized normal state should have a smaller $T_c$ as compared to the superconductivity in an unpolarized normal state. Note that only $f$-wave spin-triplet superconductivity is allowed as the spin-singlet pairing is completely suppressed in a spin-polarized normal state. In Fig.~\ref{Fig:MuStar}(b), we plot the effective pairing $g^*$ in the presence of Coulomb repulsion as a function of doping for $\Delta_1=40-80$ meV, and demonstrate that $g^*$ remains positive in the vicinity of VHS. 
	Within the mean-field approximation, we find that superconductivity exists near VHS in hole doping, and the highest $T_c$ ($\sim 20$ mK) depends on $\epsilon$ and $d$ as shown in Fig.~\ref{Fig:BBG}(b). In the electron-doped regime, we also find superconductivity with the highest $T_c\sim 0.5$ K close to VHS, and $T_c$ is insensitive to $\epsilon$ and $d$, suggesting strong screening of Coulomb potential.

	\textit{Discussion.---} The recent BBG experiment \cite{Zhou2021_BBG} reported the existence of spin-triplet superconductivity near $n_e=0.57\times 10^{12}$ cm$^{-2}$ in the presence of a finite in-plane magnetic field, while no superconducting phase is found in the absence of a magnetic field for $|n_e|<0.8\times 10^{12}$ cm$^{-2}$ in the hole doping. The experiment also found multiple symmetry-breaking phases. We argue below that our proposed electron-acoustic-phonon mechanism can explain the BBG superconductivity phenomenology \cite{Zhou2021_BBG}.
	
	In the experiment \cite{Zhou2021_BBG}, a low-temperature competing ordered state (with an insulator-like nonlinear resistivity) exists at $n_e=-0.59\times 10^{12}$ cm$^{-2}$ in the absence of a magnetic field. A sufficiently large magnetic field suppresses the insulator-like state and creates a partially ``isospin'' polarized metal with a finite spin polarization. Based on our theory, $f$-wave spin-triplet superconductivity from a spin-polarized valley-polarized metal happens only near VHS. For $\Delta_1=50$ meV and a spin-polarized metal, VHS is at $n_e=-0.6\times 10^{12}$ cm$^{-2}$ (half of the $n_e$ in an unpolarized metal) which is in good agreement with the experimentally found $n_e=-0.57\times 10^{12}$ cm$^{-2}$ where superconductivity is observed \cite{Zhou2021_BBG}. Assuming that the metallic state near VHS is nearly spin-polarized by the in-plane magnetic field, we obtain the highest $T_c\sim 20$ mK which is in quantitative agreement with the experimentally extracted $T_c\approx 26$ mK. Our explanation is that the superconductivity is preempted by the insulator-like state near VHS, and superconductivity becomes the dominant instability after the competing insulating state is eliminated by an in-plane magnetic field.
	In Fig.~\ref{Fig:BBG}(b), we predict $T_c$ as a function of the half gate distance $d$ for a few representative values of dielectric constant $\epsilon$, which should be tested experimentally.
	
	The absence of superconductivity without a magnetic field for $|n_e|<0.8\times 10^{12}$ cm$^{-2}$ in the hole doped regime is also consistent with our theory. Again, the interplay between acoustic-phonon attraction and Coulomb repulsion results in observable superconductivity only in the vicinity of VHS. For an unpolarized metal, the positions of VHS [peaks in Fig.~\ref{Fig:DOS_Tc}(a)] for $|\Delta_1|>40$ meV are outside the parameter range in the experiments. We predict the existence of superconductivity for larger doping as shown in Fig.~\ref{Fig:BBG}(c). Note that superconductivity might be superseded by interaction-induced symmetry-breaking phases, which we do not consider in our theory. We also predict observable superconductivity in the electron-doped side for both an unpolarized metal and for a spin-polarized half metal. 
	
	Since electron-acoustic-phonon coupling is essential, one should expect a linear-in-$T$ resistivity for $T>T_{\text{onset}}$ and a $T^4$ resistivity for $T<T_{\text{onset}}$ \cite{Hwang2008,Min2011}. Based on Ref.~\cite{Min2011}, we estimate that $T_{\text{onset}}$ is around 10-20K, which is beyond the highest temperature measured in the BBG experiment \cite{Zhou2021_BBG}. We anticipate that a linear-in-$T$ resistivity manifests for $T>20$K near the same doping density where superconductivity is observed.
	
	Finally, we comment on the universal theory for superconductivity in graphene-based materials. Since the superconductivity phenomenology in RTG \cite{Chou2021_RTG_SC} and in BBG, as well as in twisted bilayer graphene \cite{Wu2019_phonon}, can be explained by acoustic phonons, it is likely, based on simply Occam's razor, that an electron-acoustic-phonon mechanism accounts for superconductivity in all graphene-based materials. We note that interaction effects are still important as they can induce symmetry-breaking correlated phases, suppressing and preempting superconductivity. Moreover, subleading pairing mechanisms either from optical phonons or from other origins (e.g., isospin fluctuations) may enhance $T_c$, lifting the degeneracy between $s$-wave spin-singlet and $f$-wave spin-triplet pairings. Our work presents the first qualitative and semi-quantitative theory for BBG superconductivity in reasonable agreement with experiment, and we speculate that acoustic phonons are responsible for superconductivity in graphene-based materials in general.
	
	\begin{acknowledgments}
		\textit{Acknowledgments.--} We are grateful to Andrea Young for sharing with us unpublished data and for sending us a preprint of Ref. \cite{Zhou2021_BBG} prior to posting. 
		This work is supported by the Laboratory for Physical Sciences. It is also partially funded by NSF-PFC at JQI (Y.-Z.C.) and by NSF DMR1555135 (J.D.S.).
	\end{acknowledgments}	
	
	

	
	\newpage \clearpage 
	
	\onecolumngrid
	
	\begin{center}
		{\large
			Acoustic-phonon-mediated superconductivity in Bernal bilayer graphene
			\vspace{4pt}
			\\
			SUPPLEMENTAL MATERIAL
		}
	\end{center}

	\setcounter{figure}{0}
	\renewcommand{\thefigure}{S\arabic{figure}}
	\setcounter{equation}{0}
	\renewcommand{\theequation}{S\arabic{equation}}
	
	In this supplemental material, we provide some technical details for the main results in the main text.

	\section{Single-particle Hamiltonian and band structure}
	
	We adopt the $\vex{k}\cdot\vex{p}$ Hamiltonian from Ref.~\cite{Jung2014}. The $\hat{h}_{\tau}(\vex{k})$ in the main text is given by
	\begin{align}
		\hat{h}_{\tau}(\vex{k})
		=\left[\begin{array}{cccc}
			-\Delta_1 & v_0\Pi^{\dagger}(\vex{k}) & -v_4\Pi^{\dagger}(\vex{k}) & -v_3\Pi(\vex{k}) \\[2mm]
			v_0\Pi(\vex{k}) & \Delta'-\Delta_1 & t_1 & -v_4\Pi^{\dagger}(\vex{k}) \\[2mm]
			-v_4\Pi(\vex{k})& t_1 & \Delta'+\Delta_1 & v_0\Pi^{\dagger}(\vex{k}) \\[2mm]
			-v_3\Pi^{\dagger}(\vex{k})& -v_4\Pi(\vex{k}) & v_0\Pi(\vex{k})  & \Delta_1
		\end{array}
		\right],
	\end{align}
	where $\Pi(\vex{k})=\tau k_xa_0+ik_ya_0$, $a_0$ is the lattice constant of graphene and $\Delta_1$ encodes the electric potential difference from the displacement field. The other parameters are given by $v_0=2.261$ eV, $v_3=0.245$ eV, $v_4=0.12$ eV, $t_1=0.361$ eV, and $\Delta'=0.015$ eV. The basis of the matrix is (1A,1B,2A,2B).\\
	
	The low-energy bands of BBG have large probability on the A sites of the top layer (1A) and B sites of the bottom layer (2B). This property arises from the interlayer nearest-neighbor tunnelings which tend to form dimerized bonds between 1B and 2A sites. To gain an intuitive understanding, one can use a $2\times 2$ Dirac Hamiltonian per spin per valley to describe the low-energy theory of BBG.\\
	
\begin{figure}[h]
		\includegraphics[width=0.45\textwidth]{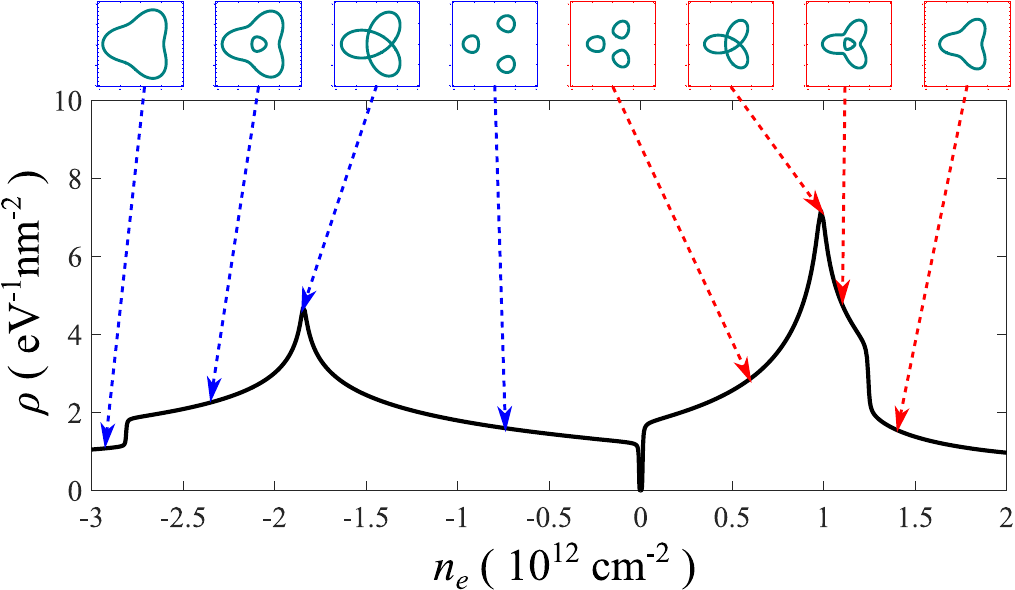}
		\caption{Density of states and evolution of Fermi surfaces. We construct DOS with a $10000\times 10000$ momentum grid with momentum cutoff $\Lambda\sim 0.15 a_0^{-1}$, where $a_0$ is the lattice constant of graphene. The shapes of Fermi surfaces and the corresponding carrier densities are labeled in both the electron-doped and hole-doped regimes.}
		\label{Fig:FS}
	\end{figure}

	In this work, we focus on $\Delta_1=40-80$ meV, corresponding to the experimentally relevant regime. As illustrated in Fig.~\ref{Fig:FS}, BBG manifests multiple transitions in the band structure. For $|\Delta_1|>0.038$ meV, there exists a region where annular Fermi surfaces are realized in the electron doping. The band structure with $|\Delta_1|=38$ meV realizes a higher order VHS in electron doping. The same features also happen in hole doping with a smaller critical value $|\Delta_1|=10$ meV.

	\section{Fermi velocity}
	
\begin{figure}[h]
		\includegraphics[width=0.45\textwidth]{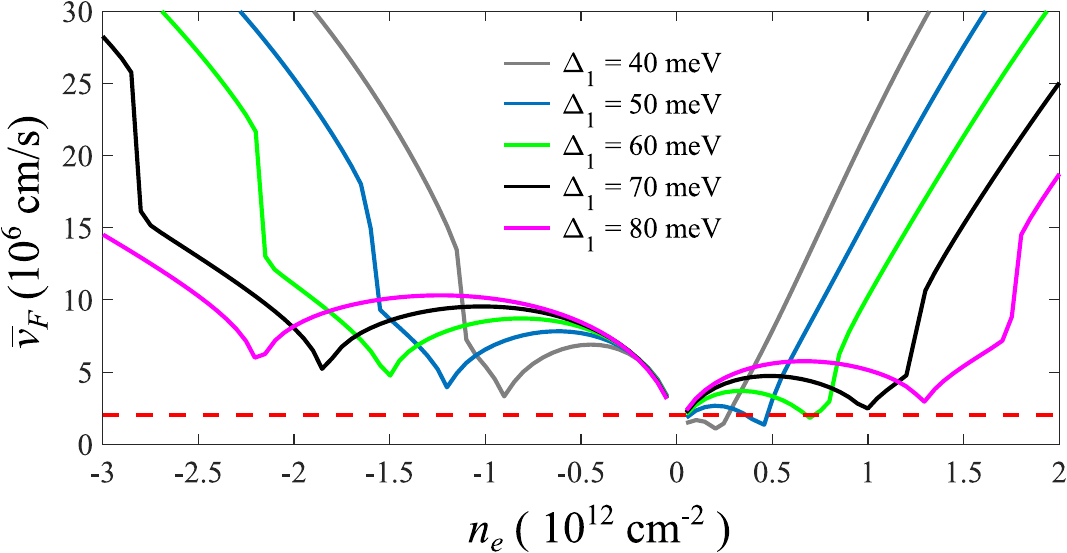}
		\caption{The average Fermi velocity ($\bar{v}_F$) as a function of doping for $\Delta_1=40-80$ meV.}
		\label{Fig:vF}
	\end{figure}

	One issue in the BCS approximation for BBG is that the Fermi velocity ($v_F$) might not be much larger than the acoustic phonon velocity ($v_s$), implying that retardation might not take place. To analyze this, we estimate the average Fermi velocity $\bar{v}_F$ as a function of doping and density of states in the following.
	\begin{align}
		&\rho=g\frac{k_F}{2\pi\hbar \bar{v}_F}=g\frac{\sqrt{4\pi |n_e|/g}}{2\pi\hbar \bar{v}_F}\rightarrow \bar{v}_F=\frac{\sqrt{g}}{\sqrt{\pi}}\frac{\sqrt{|n_e|}}{\rho\hbar}\rightarrow \bar{v}_F=\frac{\sqrt{g}}{\sqrt{\pi}}\frac{\sqrt{|n_e| \times (10^{-12}\text{cm}^2})}{\rho\times( \text{eV.nm}^2)}\frac{1}{6.582}10^8\text{cm/s},
	\end{align}
	where $g=4$ is the degeneracy factor. Note that $\rho$ is the total density of states. Therefore,
	\begin{align}
		\bar{v}_F=\frac{\sqrt{|n_e| \times (10^{-12}\text{cm}^2})}{\rho\times( \text{eV.nm}^2)}\times1.71434\times 10^7\text{cm/s}.
	\end{align}
	In Fig.~\ref{Fig:vF}, $\bar{v}_F>v_s$ for generic doping densities, backing up that the assumption of retardation. The exceptions are (1) $n_e$ close to VHS (not resolved in our numerics) and (2) low carrier density in the electron doping. Interestingly, our results suggest that retardation remains valid for $n_e$ slightly away from VHS.

	\section{Derivation of gap equations}
	
	In this section, we focus only on the $b$th band and then drop the band index. With the mean field approximation, $\hat{H}_0+\hat{H}_{\text{BCS}}'$ becomes:
	\begin{align}
		\hat{H}_{\text{MFT}}=&\sum_{s,s'}\sum_{\vex{k}}\mathcal{C}^{\dagger}_{ss'}(\vex{k})\hat{h}_{\text{BdG},s's}(\vex{k})\mathcal{C}_{ss'}(\vex{k})+\mathcal{A}\sum_{s,s'}\sum_{\vex{k},\vex{k'}}\Delta^*_{s's}(\vex{k})\left[\left(g\right)^{-1}\right]_{\vex{k},\vex{k}'}\Delta_{s's}(\vex{k}'),
	\end{align}
	where
	\begin{align}
		\mathcal{C}^{T}_{ss'}(\vex{k})=&[c_{+ bs}(\vex{k}); c_{-bs'}^{\dagger}(-\vex{k})],\\
		\hat{h}_{\text{BdG},s's}=&\left[\begin{array}{cc}
			\mathcal{E}_{+}(\vex{k})-\mathcal{E}_F & \Delta_{s's}(\vex{k})\\[2mm]
			\Delta^*_{s's}(\vex{k}) & -\mathcal{E}_{-}(-\vex{k})+\mathcal{E}_F
		\end{array}\right],\\
		\Delta_{s's}(\vex{k}')=&\frac{1}{\mathcal{A}}\sum_{\vex{k}'}g_{\vex{k},\vex{k}'}\left\langle c_{-bs'}(-\vex{k}')c_{+bs}(\vex{k}')\right\rangle.
	\end{align}

	In the imaginary-time path integral, we integrate out the fermions and derive the effective action given by
	\begin{align}
		\mathcal{S}_{\text{eff}}=&-\sum_{\omega_n,\vex{k}}\ln\left[\left(-i\omega_n+\mathcal{E}_{+}(\vex{k})-\mathcal{E}_F\right)\left(-i\omega_n-\mathcal{E}_{-}(-\vex{k})+\mathcal{E}_F\right)-|\Delta(\vex{k})|^2\right]+\mathcal{A}\beta\sum_{\vex{k},\vex{k'}}\Delta^*(\vex{k})\left(g^{-1}\right)_{\vex{k},\vex{k}'}\Delta(\vex{k}'),
	\end{align}
	where we have suppressed the spin indices and band index for simplicity.
	Near the transition temperature, we assume the $\Delta(\vex{k})$ is infinitesimal and expand the logarithm. As a result, we obtain the Landau free energy density as follows:
	\begin{align}
		\mathcal{F}=&\frac{\mathcal{S}_{\text{eff}}}{\beta \mathcal{A}}\approx\text{const}+\frac{1}{\beta\mathcal{A}}\sum_{\omega_n,\vex{k}}\frac{|\Delta(\vex{k})|^2}{\left(-i\omega_n+\mathcal{E}_{+}(\vex{k})-\mathcal{E}_F\right)\left(-i\omega_n-\mathcal{E}_{-}(-\vex{k})+\mathcal{E}_F\right)}+\sum_{\vex{k},\vex{k'}}\Delta^*(\vex{k})\left(g^{-1}\right)_{\vex{k},\vex{k}'}\Delta(\vex{k}')+\mathcal{O}(|\Delta(\vex{k})|^4)\\
		=&\text{const}-\frac{1}{\mathcal{A}}\sum_{\vex{k}}\frac{1-2f(\mathcal{E}_+(\vex{k})-\mathcal{E}_F)}{2\left[\mathcal{E}_+(\vex{k})-\mathcal{E}_F\right]}|\Delta(\vex{k})|^2+\sum_{\vex{k},\vex{k'}}\Delta^*(\vex{k})\left(g^{-1}\right)_{\vex{k},\vex{k}'}\Delta(\vex{k}')+\mathcal{O}(|\Delta(\vex{k})|^4),
	\end{align}
	where $f(x)$ is the Fermi distribution function. We have used the spinless time-reversal symmetry yielding $\mathcal{E}_{+}(\vex{k})=\mathcal{E}_{-}(-\vex{k})$.
	The linearized gap equation can be obtained by differentiating $\Delta^*(\vex{k})$ on $\mathcal{F}$,
	\begin{align}
		&\frac{\delta \mathcal{F}}{\delta \Delta^*(\vex{k})}=0=-\frac{1}{\mathcal{A}}\frac{1-2f(\mathcal{E}_+(\vex{k})-\mathcal{E}_F)}{2\left[\mathcal{E}_+(\vex{k})-\mathcal{E}_F\right]}\Delta(\vex{k})+\sum_{\vex{k}'}\left(g^{-1}\right)_{\vex{k},\vex{k}'}\Delta(\vex{k}').
	\end{align}
	Note that the momentum indices can be viewed as matrix indices. After some algebraic manipulation, we obtain
	\begin{align}
		\Delta(\vex{k})=&\sum_{\vex{k}'}\chi_{\vex{k},\vex{k}'}\Delta(\vex{k}'),\\
		\chi_{\vex{k},\vex{k}'}=&\frac{g_{\vex{k},\vex{k}'}}{\mathcal{A}}\frac{1-2f(\mathcal{E}_+(\vex{k}')-\mathcal{E}_F)}{2\left[\mathcal{E}_+(\vex{k}')-\mathcal{E}_F\right]}=\frac{g_{\vex{k},\vex{k}'}}{\mathcal{A}}\frac{\tanh\left[(\mathcal{E}_+(\vex{k}')-\mathcal{E}_F)/(2k_BT)\right]}{2(\mathcal{E}_+(\vex{k}')-\mathcal{E}_F)},
	\end{align}
	where $T$ is the temperature.
	$\chi_{\vex{k},\vex{k}'}$ can be viewed as a matrix with the discrete momentum $\vex{k}$ and $\vex{k}'$ being the matrix indices. The transition temperature is obtained when $\chi_{\vex{k},\vex{k}'}$ yields a maximal egienvalue $1$.

	\section{Parameters in numerical calculations}

	To choose the appropriate grid spacing and energy cutoff, we need to make sure that the momentum grid spacing is small enough to capture the intricate details in the Fermi surface near Van Hove singularity, and we have to make sure that the cutoff is large enough. We have tested a few different momentum grid spacing and energy cutoff in our calculations. Specifically, we tested momentum grid spacings $\Delta k\approx$ $2\times 10^{-3}a_0^{-1}$, $1\times 10^{-3}a_0^{-1}$, and $5\times 10^{-4}a_0^{-1}$ with $5000$, $10000$, and $20000$ low-energy levels, respectively. The $T_c$ versus $n_e$ plots are almost identical, suggesting that the results are converged, so we decide to use $2\times 10^{-3}a_0^{-1}$ and $5000$ energy levels. We also tested the cutoff for energy levels and found that the results remain the same with more levels included.

	\section{Coulomb potential}
	
	The Coulomb potential in the main text corresponds to
	\begin{align}
		V_C(\vex{q})=&\frac{2\pi e^2}{\epsilon |\vex{q}|}\tanh(|\vex{q}|d)=2\pi\times (1.44\text{eV.nm})\times\frac{a_0}{|\vex{q}|a_0}\times\tanh\left(|\vex{q}|a_0\times\frac{d}{a_0}\right)=\frac{2.23}{\epsilon}\frac{\tanh\left(|\vex{q}|a_0\times\frac{d}{a_0}\right)}{|\vex{q}|a_0} \text{eV.nm}^2,
	\end{align}
	where $d$ is half gate distance and $a_0$ is the lattice constant of graphene. The $T_c$ results are mostly insensitive to $d$ and $\epsilon$ except for the hole-doped spin-triplet superconductivity emerging from a spin polarized metal.

	\section{Derivation of $\Gamma$}

	The self-consistent ladder Dyson equation in Fig.~3 of main text corresponds to an algebraic equation as follows
	\begin{align}\label{Eq:DysonE}
		-\tilde{V}'(\vex{k}'-\vex{k})=-V(\vex{k}'-\vex{k})+\frac{1}{\beta \mathcal{A}}\sum_{\omega_c<|\omega_n|<\Lambda}\sum_{\vex{p},|\tilde{\mathcal{E}}_{+}(\vex{p})|<\Lambda}V(\vex{p}-\vex{k})\frac{1}{\omega_n^2+\tilde{\mathcal{E}}_{+}^2(\vex{p})}\tilde{V}(\vex{k}'-\vex{p}),
	\end{align}
	where $\tilde{\mathcal{E}}_{+}(\vex{p})=\mathcal{E}_+(\vex{p})-\mathcal{E}_F$, $\tilde{V}$ is the renormalized interaction, and $V$ is the bare interaction. To simplify the calculations, we drop the momentum dependence and use $U_0(\mathcal{E}_F)\equiv V_{\text{TF}}(k_F;\mathcal{E}_F)$ with $k_F$ estimated by $\sqrt{4\pi |n_e|/g}$ where $g=4$ is the degeneracy factor. Solving Eq.~(\ref{Eq:DysonE}), we obtain
	\begin{align}
		U_R(\mathcal{E}_F)=\frac{U_0(\mathcal{E}_F)}{1+U_0(\mathcal{E}_F)\Gamma(\mathcal{E}_F;\omega_c;\Lambda)},
	\end{align}
	where
	\begin{align}
		\Gamma(\mathcal{E}_F;\omega_c;\Lambda)=\frac{1}{\beta\mathcal{A}}\sum_{\omega_c<|\omega_n|<\Lambda}\sum_{\vex{p},|\tilde{\mathcal{E}}_{\vex{p}}|<\Lambda}\frac{1}{\omega_n^2+\tilde{\mathcal{E}}_{\vex{p}}^2}.
	\end{align}
	
	To evaluate $\Gamma(\mathcal{E}_F;\omega_c;\Lambda)$, we assume the zero temperature limit (i.e., $\beta\rightarrow \infty$) and derive
	\begin{align}
		\Gamma(\mathcal{E}_F;\omega_c;\Lambda)=\frac{2}{\mathcal{A}}\sum_{\vex{p},|\tilde{\mathcal{E}_{\vex{p}}}|<\Lambda}\int_{\omega_c}^{\Lambda}\frac{d\omega}{2\pi}\frac{1}{\omega^2+\tilde{\mathcal{E}}_{\vex{p}}^2}=\frac{1}{\mathcal{A}}\sum_{\vex{p},|\tilde{\mathcal{E}_{\vex{p}}}|<\Lambda}\frac{1}{\pi|\tilde{\mathcal{E}_{\vex{p}}}|}\left[-\tan^{-1}\left(\frac{|\tilde{\mathcal{E}}_{\vex{p}}|}{\Lambda}\right)+\tan^{-1}\left(\frac{|\tilde{\mathcal{E}}_{\vex{p}}|}{\omega_c}\right)\right].
	\end{align}
	The above expression can be efficiently evaluated numerically, and $\Gamma(\mathcal{E}_F;\omega_c;\Lambda)$ is insensitive to finite size effect. We choose $\Lambda=100$ meV in all the calculations.
	


\end{document}